\documentclass[aps,floatfix,superscriptaddress,showpacs,twocolumn]{revtex4}

\usepackage{amsfonts}
\usepackage{graphicx,floatrow}

\usepackage{array}
\usepackage{amsthm}
\usepackage{amsmath}
\usepackage{amssymb}
\usepackage{txfonts}
\usepackage{color}
\usepackage[colorlinks,citecolor=blue]{hyperref}
\usepackage{bm}
\usepackage{amsmath}

\usepackage{floatrow}
\newfloatcommand{capbtabbox}{table}[][\FBwidth]

\let\mathbb=\varmathbb
\DeclareSymbolFont{letters}{OML}{ztmcm}{m}{it}

\begin{document}

\title{Critically-enhanced  spin-nematic squeezing and entanglement in dipolar spinor condensates}

\author{Qing-Shou Tan}
\affiliation{Key Laboratory of Hunan Province on Information Photonics and Freespace Optical Communications, Hunan Institute of Science and Technology, Yueyang 414000, China}
\affiliation{College of Physics and Electronic Engineering, Hainan Normal University,
Haikou 571158, China}
\author{Yixiao Huang}
\affiliation{School of Science, Zhejiang University of Science and Technology, Hangzhou, Zhejiang, 310023, China}

\author{Qiong-Tao Xie}
\affiliation{College of Physics and Electronic Engineering, Hainan Normal University,
Haikou 571158, China}
\author{Xiaoguang Wang}
\affiliation{Zhejiang Institute of Modern Physics, Department of Physics, Zhejiang University, Hangzhou 310027, China}

\date{\today }
\begin{abstract}
We study the quantum critical effect enhanced spin-nematic squeezing and  quantum Fisher information (QFI) in the
spin-1 dipolar atomic Bose-Einstein condensate. We show that the quantum phase transitions can improve the
squeezing and QFI  in the nearby regime of critical point, and the Heisenberg-limited  high-precision metrology can be obtained.
The different properties  of  the ground squeezing and entanglement under even and odd number of atoms  are  further analyzed,
by calculating the exact analytical expressions.We  also  demonstrate   the  squeezing and entanglement  generated by 
the spin-mixing dynamics around the phase transition point. It is shown that the steady squeezing and entanglement can be obtained,
and the  Bogoliubov approximation can well describe the dynamics of spin-nematic squeezed vacuum state.
 \end{abstract}


\maketitle

\section{introduction}
Spin squeezing has attracted much attention  in precision metrology since it was first established by   Kitagawa and Ueda~\cite{Kitagawa}.
 In the past two decades, spin-squeezed states have been widely  used
in   high-precision  measurements  to beat the standard quantum limit (SQL)~\cite{ Wineland,Wineland2,ma,Pezze,Cronin, Sau, Vitagliano} which is
 the best  estimation limit of separable states with $N$ particles and  scales like $1/\sqrt{N}$.
In Ref.~\cite{Kitagawa},  two different mechanisms were proposed to generate spin-squeezed states:  one-axis twisting (OAT) and two-axis twisting (TAT). 
The precision allowed by OAT and TAT states scales with $1/N^{2/3}$ and $1/N$, respectively.
The  best precision of TAT squeezed states is known as the Heisenberg scaling. 
In experiments, the TAT squeezed states are hard to achieve, while the OAT ones  have been  applied  in Ramsey spectroscopy, 
atom interferometers and high-precision atomic clocks.

The nonlinearity of Bose-Einstein condensates (BECs) caused by  atomic collisions can create spin-squeezed states, and  is proved to be an ideal candidate 
for high resolution  quantum metrology~\cite{Gross, Riedel}.
In particular, the spinor atomic  BECs  have arisen much interest~\cite{law, chang, Kawaguchi,Stamper-Kurn,duan,you,Mustecapliogl,Kajtoch} due to
their  significant roles in studying  the quantum metrology of many-body spin systems.
Basically, these works can be sorted into two categories:  spin-1/2  and integer-spin atomic systems.
Compared with spin-1/2 atoms, whose states can be uniquely specified by different components of the total spin vector $\hat{\bf {S}}=(\hat{S}_x, \hat{S}_y, \hat{S}_z)$,
 spin-1 atoms  require additional spinor degrees of freedom to describe, associated with the quadrupole or nematic tensor operator,
 $\hat{Q}_{ij}$ $({i,j}\in{x,y,z})$ \cite{Hamley,Gerving,Hoang,Huang,Masson,Masson2,Niezgoda}.
These  additional  degrees of freedom concomitantly offer more degrees of freedom to  squeezing and entanglement.
 Recently, the spin-nematic squeezing  was observed in  experiment  by the nonlinear collisional dynamics of spinor  BEC,
  and the  squeezing can  be improved on the SQL by up to 8-10 dB~\cite{Hamley}.

In spinor atomic BECs, besides nonlinear collisional interactions, there is also  long-range magnetic dipole-dipole interaction (MDDI)~\cite{syi2001,syi,Stuhler,syi2,xing,Giovanazzi,Griesmaier,Puh, Zhangw,huangy}.
According to the recent experimental and theoretical observation in $^{23}$Na, $^{87}$Rb and  $^{52}$Cr atoms, the  MDDIs are indeed not negligible for these spinor condensates.
Particularly, the achievements in spinor  BECs provide a highly tunable and controllable system where the spin interactions, including the MDDI, can be accurately engineered~\cite{syi2001,syi, Giovanazzi,Chin}. 
The relative strength of the dipolar interaction and the spin exchange interaction describes a rich phase diagram~\cite{syi,syi2,huangy}.
These transitions between different phases are of interest with respect to spin squeezing and entanglement.

The present work concerns  generating highly spin-nematic squeezing and  metrologically useful  entanglement  in different phases of
spin-1 dipolar condensate including an ensemble of $N$ atoms. Both the ground states and dynamical behavior for them are considered.
 As same as  the usual spin squeezing, in spin-nematic squeezing, entanglement is  also induced in an ensemble of  atomic spins.  
Quantum Fisher information (QFI)~\cite{Helstrom,Holevo}, which  plays a central role in quantum metrology, is
able to detect useful multipartite entanglement. It is proved that QFI can perform
even better than spin squeezing parameter in the detection of non-Gaussian states~\cite{strobel}. Thus,
we can characterize the  metrologically useful entanglement  with QFI. In the SQL, the QFI $F \propto N$ is reached
when uncorrelated atoms are used, while in the Heisenberg limit (HL), $F\propto N^2$ is possible by using entangled
states.

Under our considered system, in ground state case,  there are three sharp changes for both the squeezing and QFI at the phase transition points.
More specifically,  with the change of MDDI, the QFI  ranges from  unentangled  state scaled as $N$ to highly-entangled state  scaled like  $N^2$.
It enables precision metrology to reach the Heisenberg scalar.
The optimal squeezing, similar to TAT $\propto 1/N$,  can occur nearby the regimes of vanished MDDI, but at where the behavior  is quite different for even and odd $N$.
In dynamics case,  we focus on the  steady squeezing nearby the critical point at which the spin transfer rates are very low.
In this case it is possible to obtain analytical prediction for spin-nematic squeezed vacuum state and QFI
by adopting Bogoliubov approximation. We also show that  the analytical results are in well agreement  with the numerical calculations.
Our results shed new light on obtaining metrologically useful  entanglement to improve the precision of quantum metrology using spinor BECs.

This  work is organized as follows. In Sec.~\ref{model}, we introduce the physical model of a spin-1 dipolar  condensates and present the spin-nematic squeezing parameter and QFI.
 In Secs.~\ref{ground} and~\ref{dynamics}, we study the  critical effect enhanced spin-nematic squeezing and QFI in the cases of ground states and dynamics, respectively.  
 Finally, a conclusion will be presented in Sec.~\ref{conclusion}.

\section{FORMULATION}\label{model}
\subsection{Model}
We consider a trapped gas of $N$ bosonic atoms with hyperfine spin
$f=1$. Atoms interact via s-wave collisions and dipolar interaction.
Assuming all spin components share a common spatial mode $\phi(r)$,
under the single-mode approximation, the total spin-dependent Hamiltonian
reads \cite{syi,syi2,xing}
\begin{eqnarray}\label{ham1}
\hat{H}=(c_{2}'-c_{d}')\hat{\bf S}^{2}+3c_{d}' \hat{S}_{z}^{2}+3c_{d}' \hat{a}_{0}^{\dagger}\hat{a}_{0}.
\end{eqnarray}
The total many-body angular momentum operator is  $\hat{\bf S}=\sum_{\alpha,\beta}\hat{a}_{\alpha}{ \bm F}_{\alpha\beta}\hat{a}_{\beta}$
 $(\alpha,\beta \in 0,\pm1 )$ with $\bm F$
being the spin-1 matrices and $\hat{a}_{\alpha}$ the annihilation
operator associated with the condensate mode,  and the magnetization is defined as $S_z=\hat{a}_1^{\dagger}\hat{a}_1-\hat{a}_{-1}^{\dagger}\hat{a}_{-1}$.
The rescaled collisional
and dipolar interaction strengths, respectively, are given by $c_{2}'=(c_{2}/2)\int dr|\phi(r)|^{4}$
and $c_{d}'=(c_{d}/4)\int drdr'|\phi(r)|^{2}|\phi(r')|^{2}(1-3\cos^{2}\theta_{e})/|\vec{r}-\vec{r'}|^{3}$
with $\theta_{e}$ being the polar angle of $(\vec{r}-\vec{r'})$.
Here $c_{2}=4\pi\hbar^{2}(a_{2}-a_{0})/(3M)$ with $M$ being the
mass of the atom and $a_{0,2}$ the $s$-wave scattering length for
two spin-1 atoms in the symmetric channel of the total spin 0 and
2, respectively. The strength of the MDDI is given by $c_{d}=\mu_{0}g_{F}^{2}\mu_{B}^{2}/4\pi$  with $\mu_{B}$
the Bohr magneton, and $g_{F}$ the Land\'e $g$-factor.

To proceed, it is convenient to rescale the Hamiltonian by using $|c_{2}'|$ as
the energy unit, which yields the dimensionless Hamiltonian
\begin{eqnarray}\label{ham2}
\hat{H}/|c_{2}'|=(\pm1-c)\hat{\bf S}^{2}+3c\hat{S}_{z}^{2}+3c \hat{a}_{0}^{\dagger}\hat{a}_{0}.
\end{eqnarray}
The sign of $ ``+" $ $(``-")$  corresponding to $c_2'>0$ $ (c_2'<0)$, which is
determined
by the type of atoms: for $c_2' <0$ (as for $^{87}$Rb) the interaction
term favors the ferro-magnetic phase; whereas for $c_2'>0$ (as for
$^{23}$Na) the antiferro-magnetic phase minimizes the interaction
energy.
Here $c\equiv c_{d}'/|c_{2}'|$ is the relative strength of the dipolar interaction with respect to the spin exchange interaction, and is
 treated as a control parameter.
Fortunately,  the sign and magnitude of the dipolar interaction
strength $c_{d}'$ can be tuned via modifying the trapping geometry~\cite{syi} or a quick rotating orienting field~\cite{Giovanazzi},
and the contact interaction strength $c_{2}'$ is also tunable via
Feshbach resonance. 
Without loss generally,  throughout this paper we focus on the case of antiferromagnetic  Bose-Einstein condensate, such that $c_2'>0$.

Due to the dipolar interaction more new quantum phases can be found by  tuning the values of $c$~\cite{syi}.
The $c$-dependence  ground state of Hamiltonian~(\ref{ham2})  can be found by minimizing $\langle{\hat{H}}\rangle$
in the $|S,m\rangle$ basis,
which is defined by
\begin{eqnarray}
\hat{\bf S}^2|S,m\rangle=s(s+1)|S,m\rangle, \hspace{0.2cm} \hat{S}_z|S,m\rangle=m |S,m\rangle,
\end{eqnarray}
where $m=0,\pm1,...\pm S$.  For a given total number of atoms $N$, the allowable values of $S$ are $S=0,2,4,...,N$.
for even $N$, and $S=1,3,5,...,N$ for odd $N$.

\subsection{Spin-nematic squeezing parameter and quantum Fisher information}

In the case of spin-1 atomic Bose-Einstein condensates, the multipolar
moments can be specified in terms of both the spin vector $\hat{S}_{i}$
and nematic tensor $\hat{Q}_{ij}$ $({i,j}\in{x,y,z})$  which constitute SU(3) Lie algebra.
Based on the definition of the operator $\hat{Q}_{ij}$~\cite{Hamley,Gerving,Hoang,Huang},
\begin{eqnarray}\label{Qij}
\hat{Q}_{yz} & = & \frac{i}{\sqrt{2}}\left(-\hat{a}_{1}^{\dagger}\hat{a}_{0}+\hat{a}_{0}^{\dagger}\hat{a}_{-1}+\hat{a}_{0}^{\dagger}\hat{a}_{1}-\hat{a}_{-1}^{\dagger}\hat{a}_{0}\right), \nonumber\\
\hat{Q}_{xz} & = & \frac{1}{\sqrt{2}}\left(\hat{a}_{1}^{\dagger}\hat{a}_{0}-\hat{a}_{0}^{\dagger}\hat{a}_{-1}+\hat{a}_{0}^{\dagger}\hat{a}_{1}-\hat{a}_{-1}^{\dagger}\hat{a}_{0}\right),\nonumber\\
\hat{Q}_{xx} & = & \frac{2}{3}\hat{a}_{0}^{\dagger}\hat{a}_{0}-\frac{1}{3}\hat{a}_{1}^{\dagger}\hat{a}_{1}-\frac{1}{3}\hat{a}_{-1}^{\dagger}\hat{a}_{-1}+\hat{a}_{1}^{\dagger}\hat{a}_{-1}+\hat{a}_{-1}^{\dagger}\hat{a}_{1},\nonumber\\
\hat{Q}_{yy} & = & -\frac{1}{3}\hat{a}_{1}^{\dagger}\hat{a}_{1}+\frac{2}{3}\hat{a}_{0}^{\dagger}\hat{a}_{0}-\frac{1}{3}\hat{a}_{-1}^{\dagger}\hat{a}_{-1}-\hat{a}_{1}^{\dagger}\hat{a}_{-1}-\hat{a}_{-1}^{\dagger}\hat{a}_{1},\nonumber\\
\hat{Q}_{zz} & = & \frac{2}{3}\hat{a}_{1}^{\dagger}\hat{a}_{1}-\frac{4}{3}\hat{a}_{0}^{\dagger}\hat{a}_{0}+\frac{2}{3}\hat{a}_{-1}^{\dagger}\hat{a}_{-1}, \nonumber
\end{eqnarray}
 there are two different spin-nematic squeezing parameters in the SU(2) subspaces,
$\{\hat{S}_{x},\hat{Q}_{yz},\hat{Q}_{zz}- \hat{Q}_{yy}\}$
and $\{\hat{S}_{y},\hat{Q}_{xz},\hat{Q}_{xx}- \hat{Q}_{zz}\}$, which
are defined by~\cite{Hamley,Huang}
\begin{equation}\label{xi1}
\xi_{x(y)}^{2}=\frac{2\langle[\Delta(S_{x(y)} \cos\varphi +Q_{yz(xz)} \sin\varphi)]^{2}\rangle_{{\rm min}}}{|\langle \hat{Q}_{zz}- \hat{Q}_{yy(xx)}\rangle|},
\end{equation}
where the minimization is over all the quadrature angle $\varphi$. A state is spin-nematic squeezed if $\xi_{x(y)}^{2}<1$.

Below, we focus on the squeezing in the $\{S_{x},Q_{yz},Q_{+}\}$
subspace with  $\hat{Q}_{+}=\hat{Q}_{zz}- \hat{Q}_{yy}$. 
 The  spin-nematic squeezing parameter may be reduced  as
\begin{eqnarray}\label{xi2}
 \xi^2_x =\frac{ A-\sqrt{ B^2+C^2}} {|\langle Q_+ \rangle|}
 \end{eqnarray}
by finding the optimal squeezing angle
\begin{eqnarray*}
\varphi_{{\rm opt}} & =\begin{cases}
\frac{1}{2}\arccos\left(\frac{-B}{\sqrt{B^{2}+C^{2}}}\right) & B\leq0\\
\pi-\frac{1}{2}\arccos\left(\frac{-B}{\sqrt{B^{2}+C^{2}}}\right) & B>0
\end{cases},\\
\end{eqnarray*}
where we define
\begin{eqnarray}\label{ev}
A&=& \langle S_{x}^{2}+Q_{yz}^{2}\rangle ,  \hspace{0.5cm}
B= \langle S_{x}^{2}-Q_{yz}^{2}\rangle , \nonumber \\
C&=& \langle S_{x}Q_{yz}+Q_{yz}S_{x}\rangle.
\end{eqnarray}

\begin{figure}[htp]
\includegraphics[scale=0.50]{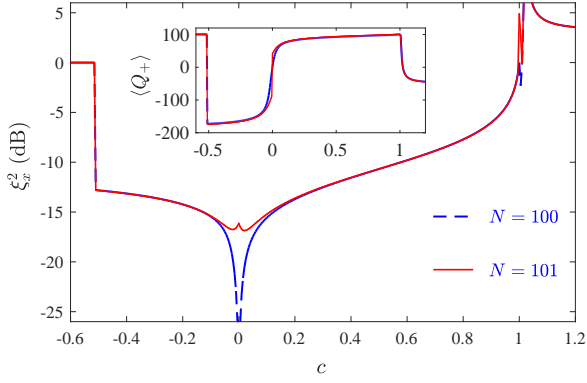}
\caption{The $c$ dependence of the spin-nematic squeezing~$\xi^2_x$ for $N=100$ and $N=101$.
The inset shows its  $\langle Q_+ \rangle$  with respect to $c$.}
\end{figure}

A wide variety of spin squeezing techniques have been used to show
sub-SQL of metrological sensitivity. To better understand the behavior
of enhanced metrological sensitivity, we  can evaluate the QFI
 in the $\hat{\Lambda}=\{\hat{S}_{x}, \hat{Q}_{yz}, \hat{Q}_{+}\}$ subspace.

According to Refs.~\cite{ma,ma2,Ferrini,Huang2,liu,Niezgoda}, the QFI $F$ with respect to measured phase $\theta$,
acquired by an SU(2) rotation on the input state $\hat{\rho}_{{\rm in}}$, can be explicitly derived as
\begin{eqnarray}\label{fi0}
F[\hat{\rho}(\theta),\hat{\Lambda}_{\vec{n}}]   = \vec{n}C\vec{n}^{T}
\end{eqnarray}
where
\begin{eqnarray}
\hat{\rho}(\theta)=\exp(-i\theta\hat{\Lambda}_{\vec{n}})\hat{\rho}_{{\rm in}}\exp(i\theta\hat{\Lambda}_{\vec{n}})
\end{eqnarray}
with $\hat{\Lambda}_{\vec{n}}=\hat{\Lambda}\cdot\vec{n}$
being the generator of rotation, and $\vec{n}$ the unit
length vector. Here the  matrix element for the symmetric matrix
$C$ is
\begin{eqnarray*}
C_{kl}=\sum_{i\neq j}\frac{(p_{i}-p_{j})^{2}}{p_{i}+p_{j}}[\left\langle i\right|\Lambda_{k}\left|j\right\rangle \left\langle j\right|\Lambda_{l}\left|i\right\rangle
 +\left\langle i\right|\Lambda_{l}\left|j\right\rangle \left\langle j\right|\Lambda_{k}\left|i\right\rangle ],
\end{eqnarray*}
where $p_{i}(|i\rangle)$ are the eigenvalues (eigenvectors) of $\hat{\rho}(\theta)$.
From Eq.~(\ref{fi0}), one finds that to get the highest possible estimation precision $\theta$, a proper direction $\vec{n}$ should be chosen for a given state,
 which maximizes the value of the QFI.
With the help of the symmetric matrix, then the maximal QFI  in the $\{\hat{S}_{x}, \hat{Q}_{yz}, \hat{Q}_{+}\}$ subspace can be obtained as
\begin{eqnarray}\label{fi}
F_{\rm max} &=& 4 \max \{ (\Delta \Lambda_{\perp})^{2}_{{\rm max}}, (\Delta Q_{+}/2)^2\}\nonumber\\
&=&   \max \left\{ 2(A+\sqrt{B^2+C^2}), (\Delta Q_{+})^2\right\},
\end{eqnarray}
where  $\langle Q_+ \rangle$ is normalized by dividing 2 since  $|\langle Q_+ \rangle|_{\rm max} =2N$.
In Eq.~(\ref{fi}), the maximal possible  value of the QFI is $F=4N^2$, which can be obtained only by the fully particle entangled states. On the
other hand, separable states can give at most $F=4N$, such as $|0,N,0\rangle$ state.
The factor 4 in the scaling of characteristic limits of
the QFI is due to  SU(3) Lie algebra~\cite{Niezgoda}.
In term of the definition in Eq.~(\ref{fi}), a state is entangled in the $\{\hat{S}_{x}, \hat{Q}_{yz}, \hat{Q}_{+}\}$ subspace if QFI $F>4N$.

In what follows, we will study the spin-nematic squeezing and QFI in the cases of ground states
and spin-mixing dynamics, respectively,  when  $c_2'>0$.

\begin{figure}[htp]
\includegraphics[scale=0.55]{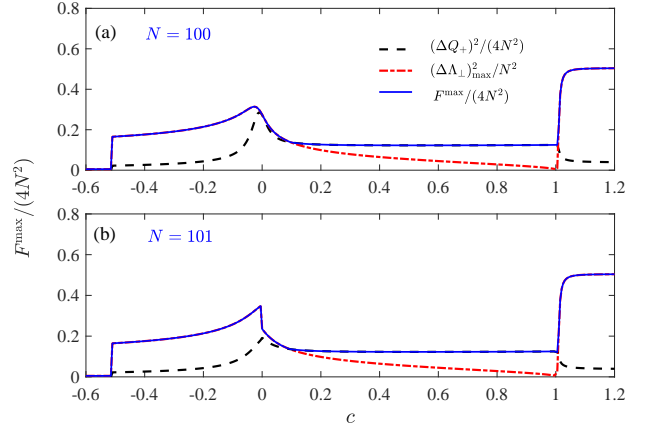}
\protect\protect\caption{ The $c$ dependence of the maximal QFI  divided by  $4N^2$ for (a) $N=100$ and (b) $N=101$.}
\end{figure}

\section{  Spin-nematic squeezing  and quantum Fisher information in ground states}\label{ground}

\begin{table*}[!htp]
\floatsetup{floatrowsep=quad,captionskip=10pt} \tabcolsep=8pt
\begin{floatrow}
\capbtabbox{
\begin{tabular}{|p{0.5cm}|p{1.1cm}|p{1.4cm}|p{1.4cm}| p{1.1cm}|}
\hline\hline
$c$              &  $<$-0.5         &  0        &  1      & $ > 1  $                            \\ \hline
$|G\rangle$     & $|N,\pm N\rangle$ &  $|0,0\rangle$&  $|\frac{N}{2},0,\frac{N}{2}\rangle$   & $\approx|N,0\rangle$    \\ \hline
$\xi^2_x$              & 1                      &  undefined    & 1            & $>1 $         \\ \hline
$F^{\rm max}$ & $2N$ & $\frac{16N(N+3)}{15}$ & $\frac{N^2}{2}+N$  & $ \approx 2N^2$  \\ \hline \hline
\end{tabular}
}{
\caption{The  ground state $|G\rangle$, the spin-nematic squeezing $\xi^2_x$, the maxima QFI $F^{\rm max}$ 
around the critical points $c<-0.5, c=0, c>=1$
for even $N$.}\label{tab:tb1}
}
\capbtabbox{
\begin{tabular}{|p{0.5cm}|p{1.1cm}|p{1.4cm}|p{1.5cm}| p{1.1cm}|}
\hline\hline
$c$           &$<$-0.5            &  0                 &  1         &   $>1$                  \\ \hline
$|G\rangle$   & $|S,\pm N\rangle$ &  $|1,0\rangle$     &  $|\frac{N+1}{2},0,\frac{N-1}{2}\rangle$    & $ \approx |N,0\rangle$   \\ \hline
$\xi^2_x$     & 1                 &  $ \frac{5}{2N+3}$ &  1         &$>1$          \\  \hline
$F^{\rm max}$ & $2N$              & $\frac{48N(N+3)-72}{35}$& $\frac{N^2-1}{2}+N$ & $ \approx 2N^2$  \\ \hline \hline
\end{tabular}
}{
 \caption{ The  ground state $|G\rangle$, the spin-nematic squeezing $\xi^2_x$, the maxima QFI $F^{\rm max}$ around 
 the critical points $c<-0.5, c=0, c>=1$
for odd $N$.}
 \label{tab:tb2}
}
\end{floatrow}
\end{table*}

Now, we will consider the spin-nematic squeezing and QFI in the case of  ground states.
Numerically, it is convenient to expand the ground state as
\begin{eqnarray}
|G \rangle = \sum_{m,k} g_{m,k}|m, k\rangle,
\end{eqnarray}
in the Fock  basis $|m,k\rangle\equiv|N_1, N_0, N_{-1}\rangle$ 
with the notations $N_{1}=k$, $N_{0}=N-2k+m$ and $N_{-1}= k-m$. Here $m = -N, -N+1,..., N$, for a given $m$, the allowable values of 
$k$ satisfy the  relation
$
{ \max}(0,m)\leq k\leq {\rm Int}\left[\frac{N+m}{2}\right],
$  where ${\rm Int}[x]$ is a function for getting the integer part of $x$.
Since Hamiltonian (\ref{ham2}) commutes with $\hat{S}_z$, the ground state must lie in certain $m$-subspace,
then the matrix elements of Hamiltonian (\ref{ham2}) become $H_{m,k,m,k'}=\langle m,k|H|m,k'\rangle$.
The amplitudes $g_{m,k}$ can be obtained just by numerically diagonalizing the Hamiltonian.
Hence the  expectation values given in Eq.~(\ref{ev})  read
\begin{eqnarray}
A= \sum_{m,k}  g_{m,k}^2 \left[(2N-4k+2m-1)(2k-m)+2N\right]
\end{eqnarray}
 and
\begin{eqnarray}
&&\sqrt{B^2+C^2}=4\sum_{m,k} |g_{m,k}g_{m,k+1}| \times \nonumber\\
&&\sqrt{(N-2k+m-1)(N-2k+m)(k+1)(k+1-m)}. \nonumber\\
\end{eqnarray}
We can also find the expectation value of $Q_{+}$
\begin{eqnarray}
\left\langle Q_{+}\right\rangle  =\sum_{m,k} g_{m,k}^{2}(6k-2N-3m),
\end{eqnarray}
as well as the corresponding fluctuation
\begin{eqnarray}
(\Delta Q_+)^2  & = &9\left[ \sum_{m,k} g_{m,k}^{2}(2k-m)^2- \left (\sum_{m,k} g_{m,k}^{2}(2k-m)\right)^2 \right]\nonumber\\
&&+ \sum_{m,k} g_{m,k}^{2} [ 2k(k+1)-m(2k+1)].
\end{eqnarray}
Substituting the above equations into Eqs.~(\ref{xi2}) and~(\ref{fi}), we can obtain the spin-nematic squeezing and QFI in the case of ground states.

Figures~1 and 2  illustrate the $c$ dependence of the spin-nematic squeezing $(10\log_{10}\xi^2_x)$ and QFI for $N=100$ (even number) and  $N=101$ (odd number),  respectively.
From  Fig.~1 and 2, we can see there are three sharp changes for both the squeezing and QFI when $c= -0.5$, $c=0$ and $c=1$.
The squeezing can be found in the region $-0.5<c<1$.
When $c<-0.5$ there is neither squeezing ($\xi^2_x=1$) nor entanglement ($F^{\rm max}=2N$), since the ground state is a Fock state with all the population in either $m_f=1$ or $-1$
 state for both the even and odd $N$.  When $c\ge1$, there is no squeezing ($\xi^2_x\ge1$) but highly-entangled states.
For instance, when $c=1$, $|G\rangle=|N/2,0,N/2\rangle$ (assuming $N$ to be even) is the Twin-Fock state~\cite{holland,hyllus,you,lucke}, 
which  is deeply entangled state in the picture of particles.
Recently, Luo \emph{et al} demonstrated near-deterministic generation of this state of ~11,000 atoms in $^{87}$Rb BEC~\cite{you}.
While for $c>1$, $|G\rangle \approx |S=N,m=0\rangle$ is the so called Dicke state~\cite{syi,duan,Wieczorek} which is a massively entangled state of all the atoms ($F^{\rm max}\approx 2N^2$).
Zhang \emph{et al}~\cite{duan} have proposed a robust method to generate this state in a spinor BEC.
From Fig.~1,  we can clearly find that the optimal squeezing  occurs around $c=0$, but the behavior  is  different for even and odd $N$  in this regime.  
Actually, for $c=0$ the spin-nematic squeezing is not well defined for even $N$, therefore, we should discuss the results
for even and odd $N$ separately, when $c=0$.

For  $c=0$  and even $N$, the ground state $|G\rangle$ is the spin-singlet state~\cite{law}
\begin{equation}
|S=0, m=0\rangle =\sum_{k=0}^{N/2} \tilde{g}_k |k, N-2k, k\rangle
\end{equation}
with $\tilde{g}_k \equiv g_{0,k}$, 
where the amplitudes obey the recursion relation
\begin{equation}
\tilde{g}_0 =\frac{1}{\sqrt{N+1}}, \hspace{0.1cm}  \tilde{g}_k=-\sqrt{\frac{N-2k+2}{N-2k+1}}\tilde{g}_{k-1}.
 \end{equation}
 After computing the recursion relation, we get
 \begin{equation}
 \tilde{g}_k=\frac{ (-1)^k}{\sqrt{N+1}} \prod_{x=0}^{k-1} \sqrt{\frac{N-2x}{N-2x-1}}.
 \end{equation}
 Spin-singlet state is a quantum superposition of a chain of Fock state in which the number of atoms in the state $m_f =\pm 1$ is equal. 
 To get some insight, we first calculate the expectation values given in Eqs.~(\ref{ev}) for this state, which yields
\begin{eqnarray}
A=\sqrt{B^2+C^2}=\frac{4N(N+3)}{15},
 \end{eqnarray}
and the expectation value of $Q_+$ is $\langle Q_+\rangle =0$.
Thus,  the spin-nematic  squeezing parameter of spin-singlet state is $0/0$ type. It is undefined at this point for even $N$, 
although the squeezing is strongest when $c \to 0$ (as shown in Fig.~1).  However, it has QFI and the optimal value is
  \begin{eqnarray}
 F^{\rm max}&=&(\Delta Q_+)^2= 4(\Delta\Lambda_{\perp})^2_{\rm max} \nonumber\\
 &=&\frac{16N(N+3)}{15},
 \end{eqnarray}
 which is the Heisenberg scalar. It indicates that the  spin-singlet state features genuine multipartite entanglement of the entire ensemble
 and will be useful for quantum metrology. This conclusion is consistent with the  earlier work reported by T\'oth  \cite{Toth}.

For  $c=0$  and odd $N$,
 the ground state  is $|G\rangle=|S=1,m=0\rangle$ which given by
\begin{eqnarray}\label{s1}
|S=1,m=0\rangle =c_0\sum_{k=0}^{n}c_k|k,N-2k,k\rangle.
\end{eqnarray}
After computing the recursion relation, the amplitudes read
\begin{eqnarray}
c_{k} = (-1)^{k}\sqrt{\frac{3(2n-2k+1)}{(k+1)}}\prod_{x=0}^{k-1}\sqrt{\frac{(x+2)(2n-2x)}{(x+1)(2n-2x-1)}},\nonumber \\
\end{eqnarray}
and the normalization constant $c_0$ is given by
\begin{eqnarray}
c_0=\left(\sum_{k=0}^{n} c_k^2 \right)^{-1/2} =\frac{1}{\sqrt{4n^{2}+8n+3}}
\end{eqnarray}
with $n=(N-1)/2$.

By substituting the ground state $|S=1,m=0\rangle$ into Eqs.~(\ref{ev}), we find
\begin{subequations}
\begin{align}
\left\langle Q_{+}\right\rangle & =  -\frac{4N+6}{5},\\
(\Delta Q_+)^2 &=\frac{128(N-1)(N+4)}{175},\\
A & =  \frac{12N^{2}+36N+12}{35},\\
\sqrt{B^{2}+C^{2}}  & =  \frac{12N^{2}+36N-48}{35}.
\end{align}
\end{subequations}
Then the value of spin-nematic squeezing is
\begin{eqnarray}
\xi^2_x=\frac{5}{2N+3}.
\end{eqnarray}
This squeezing value is quite  similar to the TAT case which  $\propto 1/N$ for $N\gg1$~\cite{Kitagawa}.
According to Eq.~(\ref{fi}),  the maximal QFI  of ground state~$|S=1,m=0\rangle$  is
\begin{eqnarray}
 F^{\rm max}=4(\Delta \Lambda_{\perp})^2_{\rm max}=\frac{48N(N+3)}{35}-\frac{72}{35},
 \end{eqnarray}
which also is the Heisenberg scalar.
To better show these  results, Table 1 (2) lists the  ground states $|G\rangle$, the spin-nematic squeezing parameter $\xi^2_x$
 and  the maximal QFI $F^{\rm max}$ around the critical points for  even (odd) $N$.

\section{  Spin-nematic squeezing  and quantum Fisher information dynamics}\label{dynamics}
We now turn to study the spin-nematic squeezing and QFI generated by the spin-mixing dynamics of the dipolar spinor condensate with even $N$.

\subsection{Numerical results}
The spin-mixing dynamics  generated squeezing and QFI
 can be studied
by numerically evolving an initial state under the total spin-dependent Hamiltonian.
Here, we consider two different initial states of the system, namely $|0,N,0\rangle$ and $|N/2,0,N/2\rangle$, and then let the states become free dynamic
evolution.   Hamiltonian (\ref{ham2}) conserves both the total particle
number $N$ and magnetization $S_z$, in general,  the evolution states have the form
\begin{equation}
|\Psi(t)=\sum_{k=0}^{N/2}g_{k}(t)|k\rangle,
\end{equation}
where $|k\rangle\equiv|k,N-2k,k\rangle$ represents the Fock state.

\begin{figure}[htp]
\includegraphics[scale=0.55]{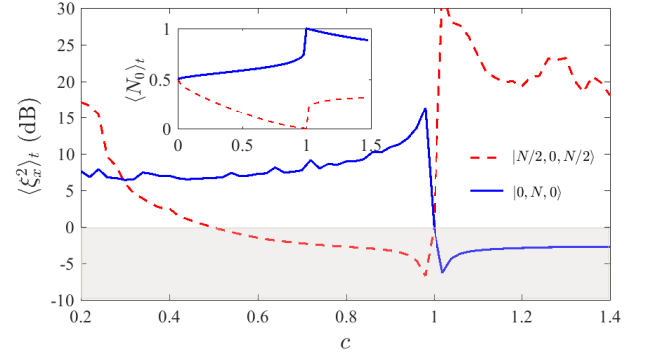}
\protect\protect\caption{The $c$ dependence of time average of $\langle{\xi_{x}^{2}}\rangle_t$ for two different initial state $|0,N,0\rangle$ and $|N/2, 0, N/2\rangle$ with $N=2000$.  
The shaded area indicates the region of squeezed. The inset shows the time average of the  population  for $m_{F}=0$
component. }
\end{figure}

\begin{figure}[htp]
\includegraphics[scale=0.53]{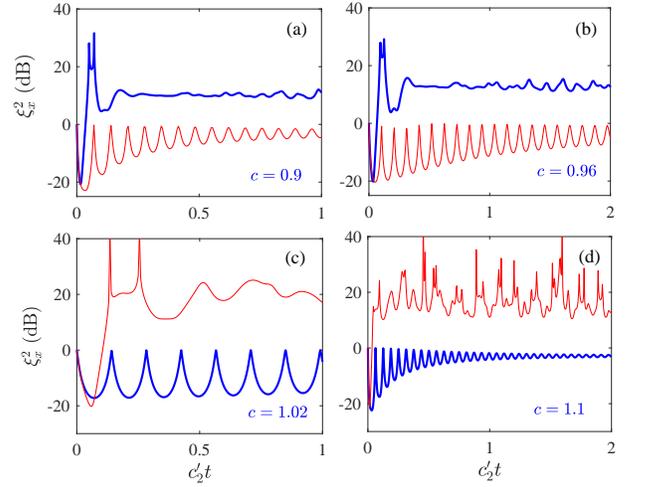}
\protect\protect\caption{Time dependence of spin-nematic squeezing parameter $\xi_{x}^{2}$
for different initial state, $|0,N,0\rangle$ (blue thick line) and $|N/2, 0, N/2\rangle$ (red thin line). Here $N=2000$.}
\end{figure}

\begin{figure}[htp]
\includegraphics[scale=0.54]{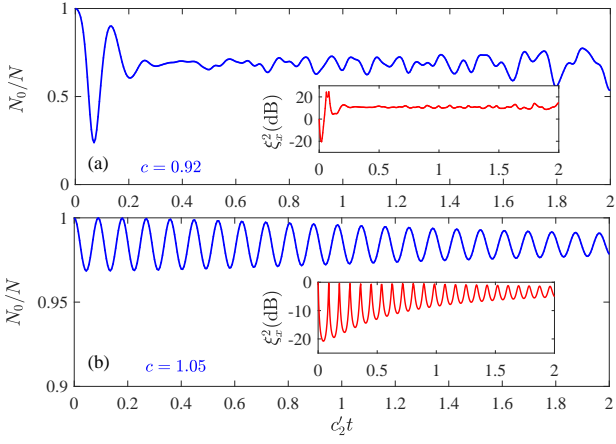}
\protect\protect\caption{
Time dependence of average number of atoms in
the $m_f=0$ mode  normalized by the total number of atoms $N$ with different $c$. The insets show the correspondant spin-nematic squeezing parameter.
The initial state of the system is $|0,N,0\rangle$ with  $N=2000$.}
\end{figure}

Here the spin-nematic squeezing parameter can be reduced to
\begin{align}\label{xxi2}
\xi_{x}^{2}  =\frac{A'-2|B'|}{|3\langle a^{\dagger}_1a_1\rangle-N|}
\end{align}
with $\langle a^{\dagger}_1a_1\rangle=\sum_{k=0}^{N/2}|g_k|^2 k$, and
\begin{subequations}
\begin{align}
A' & =  \sum_{k=0}^{N/2}|g_{k}|^{2}[(k+1)(N-2k)+(N-2k+1)k],\\
B' & =  \sum_{k=1}^{N/2}g_{k}^{*}g_{k-1}k\sqrt{(N-2k+2)(N-2k+1)}.
\end{align}
\end{subequations}
We can also find the maximal QFI as
\begin{eqnarray}
 F^{\rm max}= {\rm max} \left\{ 4(A'+2|B'|), (\Delta Q_+)^2\right\},
\end{eqnarray}
where
\begin{eqnarray}
(\Delta Q_+)^2=2\sum_k^{N/2}|g_k|^2 k(k+19) -36\left(\sum_k^{N/2}|g_k|^2 k\right)^2.
\end{eqnarray}

The spin-mixing dynamics will quickly drive the system into a quasi-steady
state~\cite{law,chang, syi}, that is  the average number of atoms in the spin components will remain unchanged for a long time.
The $c$ dependence of the quasi-steady state squeezing as well as population of the $m_{f}=0$
component for two different initial states $|0,N,0\rangle$ and $|N/2, 0, N/2\rangle$  are
plotted in Fig.~3.  As shown in Fig.~3, the quasi-steady state behavior display  sudden change when $c\to1$.
For initial state $|0,N,0\rangle$, we can get steady squeezing ($\approx$ - 5dB) in the regime of $c>1$, while the case for initial state
$|N/2, 0, N/2\rangle$ will be the opposite.
The detailed dynamical behaviors of the squeezing for these states around the critical point $c=1$ are shown in Fig.~4.
 As is shown, the spin-nematic squeezing can be improved on the SQL by up to $20$dB around the critical point before reaching the steady squeezing.

We note that the preparation of  highly entangled ideal Twin-Fock state $|N/2, 0, N/2\rangle$  may pose an experimental challenge.
Below, we  focus on the case of initial states $|0,N,0\rangle$ to understand the  steady squeezing behavior shown in Fig.~3 and 4.
In term of the spin-nematic squeezing parameter given in Eq.~(\ref{xxi2}), we can find that 
the squeezing depends on the atomic population. 
When $c\to 1$, there is essentially no population transfer from the mode $m_f=0$ to the other two modes $m_f=\pm 1$, and hence 
$\langle N_0\rangle/N \to 1$
which corresponds to squeezed vacuum for the $m_f=\pm 1$ modes~\cite{Hamley,Gerving,Hoang,Huang}.
Once $c$ deviated from 1, as evolution time is increased, the ratio $ N_0/N$ will decrease until reach the quasi-steady
state, due to the spin-mixing dynamics. In Fig.~5, we show the dynamical behavior of $N_0/N$ as well as the corresponding squeezing. 
As it is shown, when $c<1$ the average number of atoms
in the $m_f=0$ mode descend rapidly and the spin-nematic squeezed vacuum only keeps for a very short time.
While $c>1$, due to the small spin mixing parameter, $1-c$, in Hamiltonian~(\ref{ham2}), the ratio $N_0/N$  will  fall  slowly before reaching the quasi-steady,
like-damped oscillation.
Corresponding to the evolution of $N_0/N$, there is a damped oscillations  of the squeezing,
and the quasi-steady squeezing can be obtained.
However, we should point out that the steady squeezing is not a spin-nematic squeezed vacuum, 
and it is a difficult task to write out the explicit form of them.

Next, we  analytically analysis the dynamical behavior of the squeezed vacuum with Bogoliubov approximation around $c\to1$.

\subsection{Bogoliubov approximation}

We now use the Bogoliubov approximation to replace
the annihilation and creation operators for the condensate with number $N$, that is $a_{0}\approx a_{0}^{\dagger}\approx\sqrt{N}$.
Up to phase factor that we may neglect since we are later concerned
only with expectation values where the phase would cancel out. Therefore, we
can introduce the operators
\begin{eqnarray}\label{su11}
K_{x} &=& \frac{1}{2}(a_{1}^{\dagger}a_{-1}^{\dagger}+a_{1}a_{-1}), \hspace{0.5cm } K_{y} =   -\frac{i}{2}(a_{1}^{\dagger}a_{-1}^{\dagger}-a_{1}a_{-1}), \nonumber \\
K_{z} &=& \frac{1}{2}(a_{1}^{\dagger}a_{1}+a_{-1}a_{-1}^{\dagger}),
\end{eqnarray}
which belong to the SU(1,1) group and satisfy
$ [K_{x},K_{y}] = -iK_{z}, [K_{y},K_{z}] = iK_{x}$ and $[K_{z},K_{x}] =iK_{y}$.

Using the definitions in Eq.~({\ref{su11}}), the effective Hamiltonian of Eq.~(\ref{ham2}) is given by
\begin{eqnarray}
H_{{\rm eff}} & \equiv & \alpha K_{z}+\beta K_{x},
\end{eqnarray}
with $c$-dependence  parameters
\begin{eqnarray}
\alpha=2[(1-c)(2N-1)-3c], \hspace{0.3cm} \beta = 4(1-c)N.
\end{eqnarray}
In terms of the SU(1,1) operators, Eq.~(\ref{ev}) may be expressed as
\begin{eqnarray}
A=4N \langle K_z\rangle, \hspace{0.5cm}
\sqrt{B^2+C^2} = 4N |\langle K_+\rangle|,
\end{eqnarray}
where $K_{+} = K_{-} ^{\dagger}=K_{x}+iK_{y}=a^\dagger_{1}a^{\dagger}_{-1}$.
Therefore,  the spin-nematic squeezing parameter  and QFI can be reduced to
\begin{eqnarray}
\xi_{x}^{2} &=& 2\langle K_{z}\rangle-2|\langle K_{+}\rangle|, \\
F^{\rm max}&=& 8N(\langle K_z\rangle +|\langle K_+\rangle|),
\end{eqnarray}
since $\Delta Q_+ \to 0$.

To get the explicit form of both the  squeezing  and QFI, we only need to calculate the expectation values $\langle K_{z}\rangle$ and $\langle K_{+}\rangle$.
With the help of  the time evolution operator $U(t)=\exp[-iH_{{\rm eff}}t] $,
 we have
\begin{eqnarray}
\langle K_{z}\rangle & = & \langle 0,N, 0| U^{\dagger}(t) K_{z} U(t)|0,N,0\rangle \nonumber\\
 & = & \frac{\varGamma_{1}(1+\varGamma^{2})}{2(1-\varGamma^{2})^{2}},\\
|\langle K_{+}\rangle| & = & |\langle 0,N, 0| U^{\dagger}(t) K_{+} U(t)|0,N, 0\rangle|\nonumber\\
 & = & \frac{\varGamma_{1}\varGamma}{(1-\varGamma^2)^2},
 \end{eqnarray}
where
\begin{eqnarray}
\varGamma  &=&  \frac{|\beta\sin(\theta t)|}{\sqrt{\alpha^{2}-\beta^{2}\cos^{2}({\theta}t)}},  \hspace{0.2cm}
\varGamma_{1}  =  \frac{\alpha^{2}-\beta^{2}}{\alpha^{2}-\beta^{2}\cos^{2}({\theta}t)}, \nonumber\\
{\theta}  &=&  \frac{1}{2}\sqrt{\alpha^{2}-\beta^{2}}.
 \end{eqnarray}
 If $c>1$  we have $\alpha^{2}>\beta^{2}$,
a direct calculation yields
\begin{eqnarray}
\xi_{x}^{2} = \frac{\varGamma_{1}}{(1+\varGamma)^{2}}, \hspace{0.5cm}
 F^{\rm max}=\frac{4N\varGamma_{1}}{(1-\varGamma)^{2}},
\end{eqnarray}
and the optical values are given by
\begin{eqnarray}
(\xi_{x}^{2})_{\rm min} &=& \frac{|\alpha|-|\beta|}{|\alpha|+|\beta|}=\frac{2c+1}{4N(c-1)+2c+1},\\ 
F^{\rm max}(t_{\rm opt})&=& \frac{4N(|\alpha|+|\beta|)}{|\alpha|-|\beta|}=\frac{4N}{(\xi^2_x)_{\rm min}},
\end{eqnarray}
when
$t_{\rm opt} = {\pi}/{\sqrt{\alpha^{2}-\beta^{2}}}$.
The above results are valid when $c\to 1_+$, which corresponds to the spin-nematic squeezed vacuum.

Figure 6 shows the comparison  of  squeezing  and  QFI  between the exact solutions and the Bogoliubov approximation for different $c$.
As is shown,  with the increasing of $c$ the squeezing and QFI are enhanced, and the Bogoliubov approximation solutions  are
in agreement  with the exact ones when $c \to 1_{+}$. 
However,  for long time  the  squeezing is not a squeezed vacuum as shown in Fig.~4 and 5, and hence the Bogoliubov approximation will invalid.

\begin{figure}[htp]
\includegraphics[scale=0.44]{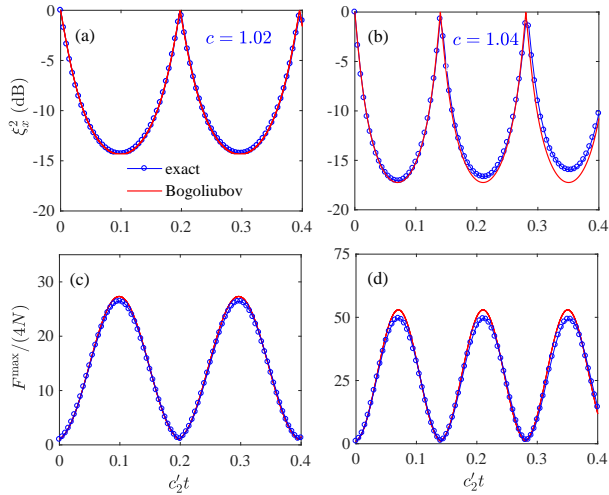}
\protect\protect\caption{Comparison of dynamical behaviors of $\xi^2_x$  (a-b)  and $F^{\rm max}/(4N)$ (c-d) for the exact numerical solution 
and Bogoliubov approximate solution with different $c$. Here $N=1000$.}
\end{figure}

\section{conclusion} \label{conclusion}
In summary, we have studied the spin-nematic squeezing and  QFI   under the ground state and spin-mixing dynamics of
a antiferromagnetic spin-1 Bose-Einstein condensate,  respectively.
We have shown that the quantum phases which depend on the relative strengths of the spin-exchange and dipolar interactions can generate
 highly entangled ground states in several limits, and  enable precision metrology to reach the HL.
We have also studied the quantum critical effect enhanced spin-nematic squeezing and  entanglement in dynamics case. It indicated that
the spin-nematic squeezing can be enhanced to $\approx$ - 20dB  before arriving the steady values $\approx$ - 5dB.
We  also  demonstrated  that the Bogoliubov approximation can well describe the dynamics of spin-nematic squeezed vacuum state.

Finally, it should be pointed out that our study here has neglected any external magnetic filed. 
The presence of external fields will affect the
orientation of the spin and hence change the phase diagram.
The effect of the magnetic field on spin-1 condensates without  MDDI is currently under study~\cite{Hoang, you,duan}. The quantum phase transition, due to the quadratic Zeeman shift,
maybe also demonstrate some similar  squeezing behavior as our case. Therefore, 
the quantum critical effect enhanced spin-nematic squeezing and entanglement  should also be expected in the case of 
spinor BEC in an external magnetic.

\begin{acknowledgments}
Q.S.T. acknowledges support from the NSFC under Grant No. 11805047 and No. 11665010, and 
Hainan Science and Technology  Plan project (Grant No. ZDKJ2019005).
Y. H.  acknowledges support from the NSFC under Grant No. 11605157. 
Q.T.X. acknowledges support from the NSFC under Grant No. 11965011.
X. W. supported by the National Natural Science Foundation of China (Grant Nos. 11875231 and 11935012), the National Key Research
and Development Program of China (Grant Nos. 2017YFA0304202 and 2017YFA0205700), and the Fundamental Research Funds
for the Central Universities (Grant No. 2018FZA3005).
\end{acknowledgments}


\begin{thebibliography}{99}

\bibitem{Kitagawa}  M. Kitagawa, M. Ueda, Squeezed spin states, Phys. Rev. A  \textbf{47}, 5138 (1993).
\bibitem{Wineland} D. J. Wineland, J.J. Bollinger, W.M. Itano, F.L. Moore, D.J. Heinzen, Spin squeezing and reduced quantum noise in spectroscopy, Phys. Rev. A \textbf{46}, R6797 (1992).
\bibitem{Wineland2}  D. J. Wineland, J.J. Bollinger, W.M. Itano, D.J. Heinzen, Squeezed atomic states and projection noise in spectroscopy, Phys. Rev. A \textbf{50}, 67(1994).

\bibitem{ma} J. Ma, X. Wang, C. P. Sun, and F. Nori, Quantum spin squeezing, Phys. Rep. \textbf{509}, 89 (2011).

\bibitem{Pezze} L. Pezz\'e, A. Smerzi, M. K. Oberthaler, R. Schmied, and P.Treutlein, Quantum metrology with nonclassical states of atomic ensembles,
Rev. Mod. Phys. \textbf{90}, 035005 (2018).

\bibitem{Cronin} A. D. Cronin, J. Schmiedmayer, D.E. Pritchard, Optics and interferometry with atoms and molecules, Rev. Mod. Phys. \textbf{81}, 1051 (2009).

 
 \bibitem{Sau} J. D. Sau, S. R. Leslie, M. L. Cohen, and D. M. Stamper-Kurn, Spin squeezing of high-spin, spatially extended quantum fields, New J. Phys. \textbf{12}, 085011 (2010).

\bibitem{Vitagliano} G. Vitagliano,  P. Hyllus, I. L. Egusquiza, and G. T\'oth,  Spin squeezing inequalities for arbitrary spin, Phys. Rev. Lett. \textbf{107}, 240502 (2011).



\bibitem{Gross} C. Gross, T. Zibold, E. Nicklas, J. \'Esteve, M.K. Oberthaler, Nonlinear atom interferometer surpasses classical precision limit, Nature \textbf{464}, 1165 (2010).

\bibitem{Riedel} M. F. Riedel, P. Bohi, Y. Li, T. W. Hansch, A. Sinatra, and P. Treutlein, Atom-chip-based generation of entanglement for quantum metrology, Nature \textbf{464}, 1170 (2010).


\bibitem{law} C. K. Law, H. Pu, and N. P. Bigelow, Quantum spins mixing in spinor Bose-Einstein condensates, Phys. Rev. Lett. \textbf{81}, 5257 (1998).

\bibitem{chang} M-S. Chang, Q. Qin, W. Zhang, and M. S. Chapman, Coherent spinor dynamics in a spin-1 Bose condensate, Nat. Phys. \textbf{1}, 111 (2005).

 \bibitem{Kawaguchi}   Y. Kawaguchi  and  M. Ueda, Spinor Bose-Einstein condensates, Phys. Rep. \textbf{520}, 253 (2012).

 \bibitem{Stamper-Kurn} D. M. Stamper-Kurn  and M. Ueda, Spinor Bose gases: Symmetries, magnetism, and quantum dynamics, Rev. Mod. Phys. \textbf{85}, 1191
(2013).


\bibitem{you} X.Y. Luo, Y. Q.  Zou, L. N. Wu,  Q. Liu,  M.  F.  Han, M. K. Tey, and  L. You, Deterministic entanglement generation from driving through quantum phase transitions, Science \textbf{355}, 620 (2017).


 \bibitem{duan} Z. Zhang  and L.-M. Duan, Generation of Massive entanglement through an adiabatic quantum phase transition
in a spinor condensate,  Phys. Rev. Lett. \textbf{111}, 180401 (2013).


\bibitem{Mustecapliogl} \"O. E. M\"ustecaplio\v{g}lu,  M. Zhang, L. You, Spin squeezing and entanglement in spinor condensates,   Phys. Rev. A \textbf{66},
033611 (2002).

\bibitem{Kajtoch} D.  Kajtoch and E. Witkowska, Spin squeezing in dipolar spinor condensates. Phys. Rev. A \textbf{93}, 023627 (2016).


\bibitem{Hamley} C. D. Hamley, C. S. Gerving, T. M. Hoang, E. M. Bookjans, and M. S. Chapman, Spin-nematic squeezed vacuum in a quantum gas, Nat. Phys. \textbf{8}, 305 (2012).

\bibitem{Gerving}  C. S. Gerving, T.M. Hoang, B.J. Land, M. Anquez, C.D. Hamley,  and M.S. Chapman,  Non-equilibrium dynamics of an unstable quantum
pendulum explored in a spin-1 Bose-Einstein condensate, Nat. Commun. \textbf{3}, 1169 (2012).

\bibitem{Hoang} T. M. Hoang, C. S. Gerving, B. J. Land, M. Anquez, C. D. Hamley, and M. S. Chapman, Dynamic stabilization of a quantum many-body spin system,
 Phys. Rev. Lett. \textbf{111}, 090403 (2013).

\bibitem{Huang} Y. Huang, H.N. Xiong, Z. Sun,  and X. Wang, Generation and storage of spin-nematic squeezing in a spinor Bose-Einstein condensate,
Phys. Rev. A \textbf{92}, 023622 (2015).

\bibitem{Masson} S. J. Masson, M. D. Barrett, and S. Parkins, Cavity QED engineering of spin dynamics and squeezing in a spinor gas, Phys. Rev. Lett. \textbf{119}, 213601 (2017).

\bibitem {Masson2} S. J. Masson and S. Parkins, Rapid production of many-body entanglement in spin-1 atoms via cavity output photon counting, Phys. Rev. Lett. \textbf{122}, 103601 (2019).

\bibitem {Niezgoda} A. Niezgoda, D. Kajtoch, and E. Witkowska, Efficient two-mode interferometers with spinor Bose-Einstein condensates, Phys. Rev. A \textbf{98}, 013610 (2018).


\bibitem{syi2001} S. Yi and L. You, Trapped condensates of atoms with dipole interactions,  Phys. Rev. A \textbf{63},  053607 (2001).
\bibitem{syi} S. Yi, L. You, and H. Pu, Quantum phases of dipolar spinor condensates, Phys. Rev. Lett. \textbf{93}, 040403 (2004).



\bibitem{syi2} S. Yi and H. Pu, Magnetization, squeezing, and entanglement in dipolar spin-1 condensates, Phys. Rev. A \textbf{73},  023602 (2006).

\bibitem{xing} H. Xing,  A. Wang, Q. S.  Tan, W. Zhang,  and S. Yi, Heisenberg-scaled magnetometer with dipolar spin-1 condensates, Phys. Rev. A \textbf{93},  043615 (2016).


\bibitem{Giovanazzi} S. Giovanazzi, A. G\"orlitz, and T. Pfau, Tuning the dipolar Interaction in quantum gases, Phys.
Rev. Lett. \textbf{89}, 130401 (2002);

\bibitem{Griesmaier}  A. Griesmaier, J. Stuhler, T. Koch,
M. Fattori, T. Pfau, and S. Giovanazzi, Comparing contact and dipolar interaction in a Bose-Einstein condensate, 
Phys. Rev. Lett. \textbf{97}, 250402 (2006).

\bibitem{Stuhler} J. Stuhler, A. Griesmaier, T. Koch, M. Fattori, T. Pfau, S. Giovanazzi, P. Pedri, and L. Santos, Observation of dipole-dipole interaction in a degenerate quantum gas,
 Phys. Rev. Lett. \textbf{95}, 150406 (2005).

\bibitem{Puh} H. Pu, W. Zhang, and P. Meystre, Ferromagnetism in a lattice of Bose-Einstein condensates, Phys. Rev. Lett. \textbf{87}, 140405
(2001).

\bibitem{Zhangw}  W. Zhang, S. Yi, M. S. Chapman, and J. Q. You, Coherent zero-field magnetization resonance in a dipolar spin-1 Bose-Einstein condensate, Phys. Rev. A \textbf{92}, 023615(2015).


\bibitem{huangy}  Y. Huang, Y. Zhang, R.  L\"u, X. Wang, and  S. Yi, Macroscopic quantum coherence in spinor condensates confined in an anisotropic potential, Phys. Rev. A \textbf{86}, 043625 (2012).

\bibitem{Chin} C. Chin, R. Grimm, P. Julienne and E. Tiesinga, Feshbach resonances in ultracold gases, Rev. Mod. Phys. \textbf{82}, 1225
(2010).



\bibitem{Helstrom}   C. W. Helstrom, Quantum detection and estimation theory
(Academic Press, New York, 1976).


\bibitem{Holevo}  A. S. Holevo, Probabilistic and statistical aspects of quantum
theory (North-Holland, Amsterdam, 1982).

\bibitem{strobel}  H. Strobel, W. Muessel, D. Linnemann, T. Zibold, D. B. Hume, L. Pezz\`e, A. Smerzi and M. K. Oberthaler, 
 Fisher information and entanglement of non-Gaussian spin states, Science \textbf{345}, 424 (2014).


\bibitem{ma2}  J. Ma, Y. Huang, X. Wang, and C. P. Sun, Quantum Fisher information of the Greenberger-Horne-Zeilinger state in decoherence channels, Phys. Rev. A \textbf{84}, 022302 (2011).

\bibitem{liu}  J. Liu, H. Yuan, X. M. Lu, X. Wang, Quantum Fisher information matrix and multiparameter estimation, J. Phys. A: Math Theoret. \textbf{53}, 023001 (2020).


\bibitem{Ferrini} G. Ferrini, D. Spehner, A. Minguzzi, and F. W. J. Hekking, Effect of phase noise on quantum correlations in Bose-Josephson junctions, Phys.
Rev. A \textbf{84}, 043628 (2011).
\bibitem{Huang2} Y. Huang, W. Zhong, Z. Sun, and X. Wang, Fisher-information manifestation of dynamical stability and transition to self-trapping for Bose-Einstein condensates, Phys. Rev. A \textbf{86},
012320 (2012).

\bibitem{holland} M. J. Holland, K. Burnett,  Interferometric detection of optical phase shifts at the Heisenberg limit, Phys. Rev. Lett. \textbf{71}, 1355 (1993).

\bibitem{hyllus} P. Hyllus, O. G\"uhne, and A. Smerzi, Not all pure entangled states are useful for sub-shot-noise interferometry,  Phys. Rev. A \textbf{82}, 012337 (2010).

\bibitem{lucke}  B. L\"ucke,  Twin matter waves for interferometry beyond the classical limit, Science \textbf{334}, 773 (2011).

\bibitem{Wieczorek}  W. Wieczorek, R. Krischek, N. Kiesel, P. Michelberger, G. T\'oth, and H. Weinfurter,  Experimental entanglement of a six-photon symmetric Dicke state,
Phys. Rev. Lett. \textbf{103}, 020504 (2009).
\bibitem{Toth} G. T\'oth, Entanglement detection in optical lattices of bosonic atoms with collective measurements,  Phys. Rev. A \textbf{69}, 052327 (2004).


\end{thebibliography}
\end{document}